\documentclass[aps,prl,superscriptaddress,twocolumn,10pt,nofootinbib,preprintnumbers,showdata]{revtex4-2}
\usepackage{bm}
\usepackage{epsfig}
\usepackage{graphics}
\usepackage{amsmath}
\usepackage[usenames, dvipsnames]{color}
\usepackage{colortbl}
\usepackage[font=small, labelfont=bf]{caption}
\begin{document}

\title{ Calculation of Thermodynamic Characteristics and \\
Sound Velocity for Two-Dimensional Yukawa Fluids Based on a\\
Two-Step Approximation for the Radial Distribution Function}
\author{ \firstname{Ilnaz~I.}~\surname{Fairushin}}
\affiliation{Department of Computational Physics, Institute of Physics, Kazan Federal
University, 420008 Kazan, Russia}
\author{ \firstname{Anatolii~V.}~\surname{Mokshin} }
\affiliation{Department of Computational Physics, Institute of Physics, Kazan Federal
University, 420008 Kazan, Russia}

\begin{abstract}
We propose a simple two-step approximation for the radial distribution function of a one-component two-dimensional Yukawa fluid. This approximation is specified by the key parameters of the system: coupling parameter and screening parameter. On the basis of this approximation, analytical expressions are obtained for the same thermodynamic quantities as internal energy, internal pressure, excess entropy in the two-particle approximation, and also longitudinal sound velocity. The theoretical results show an agreement with the results obtained in the case of a true radial distribution function.
\end{abstract}

\maketitle

\affiliation{Department of Computational Physics, Institute of Physics, Kazan Federal
University, 420008 Kazan, Russia}

\affiliation{Department of Computational Physics, Institute of Physics, Kazan Federal
University, 420008 Kazan, Russia}

The model system of charged particles with a screened Coulomb (Yukawa) interaction is widely used for modeling and predicting the physical properties of a wide variety of real fluids such as simple neutral fluids, liquid metals, colloidal solutions, microemulsions, etc.~\cite{Fortov1, Dubin}. This system represents similarly charged particles of the same mass (ions, dusty, or colloidal particles) surrounded by a background of the opposite sign, which in practice usually consists of particles of a smaller mass (electrons or ions in the case of dusty plasma or colloidal solutions) \cite{Dubin}. The interaction potential energy $u$ of a pair of charged particles on the distance $r$ in this case is written as
\begin{equation}
u(r) = \frac{(Ze)^2\exp(-r/\lambda_s)}{4\pi\varepsilon_0r}\,,
\label{yukawa}
\end{equation}
where $Z$ is the charge number, $e$ is the electron charge, $\varepsilon_0$ is the vacuum permittivity, and $\lambda_s$ is the Debye screening length associated with the presence of a neutralizing medium, which depends on the concentration and temperature of the background particles and determines how the interaction of the main particles will differ from the simple Coulomb interaction.

In the last decade, researchers have increased their attention on two-dimensional Yukawa systems \cite{Fortov1, Dubin, Klumov&KhrapakRINP, Klumov2022, Vasilieva, Kononov, Fairushin_Mol, Fairushin_JETP, Lin}. Firstly, this is due to the fact that such systems can be easily implemented in experiments. These include experiments with a monolayer of charged macroparticles in a low-temperature plasma \cite{Vasilieva, Kononov} and experiments with colloidal solutions with microsized particles \cite{Lin}. Within the framework of these experiments, it is possible to directly monitor the dynamics of individual particles, as well as to study the features of phase transitions in two-dimensional systems \cite{Vasilieva}. Second, the Yukawa interaction potential is expressed by a simple analytical expression. Thus, a system of particles interacting through a potential of the form \eqref{yukawa} can be considered as a convenient model of a one-component substance, on the example of which it is convenient to test one or another microscopic theory describing the structure or dynamics of simple substances. This is especially important for liquid-like disordered systems, which, unlike gases and crystalline bodies, are characterized by the absence of a suitable small parameter for the development of an appropriate theoretical description \cite{Hansen, MokshinTMP, Khrapak2022, Ryzhov}. The key characteristic of the liquid structure and its short-range order is the radial distribution function (RDF) $g(r)$. Knowing the RDF for systems with a known interparticle interaction potential $u(r)$, one can directly calculate such thermodynamic parameters of many-particle systems as internal energy, internal pressure, and excess entropy in the pairwise approximation \cite{HamaguchiJCP, Hartmann, KhrapakJCP, KhrapakPoP2016, Filippov, SExcess, Yurchenko}. {Note that the RDF is a special case of the radial basis functions, which are widely used in various problems of fluid mechanics \cite{Geronzi, Groth}.}

The specificity of the interparticle interaction in the case of the Yukawa system is determined by two key dimensionless parameters: coupling parameter $\Gamma$ and screening parameter $\kappa$ \cite{Fortov1, Dubin, MokshinAV}. The coupling parameter
\begin{equation}
\Gamma = \frac{(Ze)^2}{4\pi\varepsilon_0ak_BT}\
\label{gamma}
\end{equation}
is the ratio of the average potential energy of interaction (without screening) to the average energy of the particle thermal motion. In the expression (\ref{gamma}), the quantity $a = (\pi \rho)^{-1/2}$ is half the average interparticle distance or the so-called radius of the Wigner--Seitz cell, $\rho$ is the number particles per unit area of a two-dimensional system, $k_B$ is the Boltzmann constant, and $T$ is the absolute temperature of the system. The screening parameter $\kappa$ is defined as the ratio of $a$ to the Debye length $\lambda_s$:
\begin{equation}
\kappa = \frac{a}{\lambda_s}\,. \label{kappa}
\end{equation}

The time scale of the charge density fluctuations in the system is determined by the plasma frequency
\begin{equation}
\omega_p = \sqrt{\frac{Z^2e^2\rho}{2a\varepsilon_0m}}\,,
\label{wp}
\end{equation}
where $m$ is the particle mass.

In \cite{Klumov}, on the bases of the quasi-localized charge approximation (QLCA) \cite{Kalman, Golden, Donko}, simple analytical expressions are obtained that describe the dispersions of longitudinal and transverse collective excitations in a three-dimensional Yukawa fluid. This goal was achieved by the authors by using the so-called one-step {(1\rm st)} approximation for the RDF:
\begin{equation}
g^{(1\rm st)}(x)=\theta(x-x_{\rm eff})\\,
\label{RDF_1st}
\end{equation}
where $\theta(x)$ is the Heaviside function, $x_{\rm eff}$ is the effective radius of the particle, which is an input parameter in this approximation, and which is determined from the known data on the internal pressure and internal energy systems. It should be noted that the one-step approximation (\ref{RDF_1st}) actually corresponds to the function $g(x)$ of a highly rarefied gas of absolutely hard spheres of radius $x_{\rm eff}$. This approximation ignores the presence of local short-range order in liquids, which manifests itself in a characteristic maximum in the function $g(x)$.

In this paper, we propose a two-step approximation for the function $g(x)$ of a two-dimensional Yukawa system. It should be noted that, initially, this approximation was proposed for the three-dimensional system \cite{Fairushin}. Here, two key parameters of the Yukawa system are used as input parameters: $\Gamma$ and $\kappa$. Within the framework of this approximation, the internal energy, internal pressure, excess entropy in the two-particle approximation, and the longitudinal sound velocity of a two-dimensional Yukawa fluid are calculated. The ($\Gamma,\; \kappa$)-states of the Yukawa fluid will be considered, where $\Gamma$ = 20; 50; and 100 and $\kappa$ = 1; 1.5; and 2. The theoretical results are compared with the calculation results based on the true RDF obtained by us using molecular dynamics (MD) simulations. The equilibrium molecular dynamics simulations of the Yukawa fluid for $\Gamma = 20, 50,$ and $100$ and $\kappa = 1, 1.5,$ and $2$ were carried out using the computational package LAMMPS \cite{LAMMPS}. The simulation was performed for a system consisting of 2500 particles interacting through a potential~(\ref {yukawa}) in a square cell, on which periodic boundary conditions were imposed. The calculations were performed in the $NVT$ ensemble. The particle motion equations were integrated in accordance with the Verlet algorithm with a time integration step $t_{\rm step}=0.01/\omega_p$. Averaging over 10,000 time steps was used to calculate the RDF.

The two-step {(2\rm st)} approximation for the function $g(x)$ is provided as:
\begin{equation}\label{eq_gr}
g^{(2\rm st)}(x) = g_m \theta(x - x_1) \theta(x_2 - x) + \theta(x - x_2).
\end{equation}

Here, the distances $x_{1}$ and $x_{2}$ determine the position and width of the first maximum of the function $g(x)$ on the $x$ axis, and $g_m$ is the height of this maximum. The presence of the maxima corresponding to the second, third, and other co-ordinations are ignored within the framework of the two-step approximation (\ref{eq_gr}). Following \cite{Ott}, we define the distance $x_{1}$ as the size of the region of absence of interparticle correlations, which is provided by the condition $g(x_1)=0.5$. Further, the distance $x_1$ can be related to the coupling parameter $\Gamma$ and the screening parameter $\kappa$ \cite{Ott}:

\begin{equation}\label{eq_Gamma_r1}
x^2_1 = \frac{1}{b_1}\ln \frac{\Gamma - b_2(\kappa)}{b_3(\kappa)},
\end{equation}
where
\begin{equation}
b_1=2.434,\nonumber
\end{equation}
\begin{equation}
b_2(\kappa)=-5.21+6.866\kappa-2.492\kappa^2 \,,\nonumber
\end{equation}
\begin{equation}
b_3(\kappa)=0.712-0.572\kappa+0.437\kappa^2 \,. \nonumber
\end{equation}

The values of the constant coefficients in these polynomials can be determined by numerically solving a system of seven non-linear Equations (\ref{eq_Gamma_r1}) written for seven different states with $\Gamma = 20$, $50$, and $100$; $\kappa = 1$, and $2$, and also $\Gamma = 50$, $\kappa = 1.5$. In this case, the distances $x_1$ for these states were determined from the true $g(r)$ obtained from the results of our MD simulation.
The solution of the system of equations was carried out using the modified Newton method, the accuracy of the numerical solution of the system of equations was at least $99.7$ \% (this is quite sufficient for the purposes of this work).
On the other hand, the distance $x_{2}$, which determines the size of the first coordination shell within the framework of the (\ref{eq_gr}) approximation, can be found from the condition of the charge neutrality of the system under consideration, which in the two-dimensional case is written as
\begin{equation}\label{charge_n1}
\int_0^\infty[1-g(x)]xdx = \frac{1}{2}.
\end{equation}

From the expression (\ref{charge_n1}) and taking into account the relation (\ref{eq_gr}), we obtain:
\begin{equation}\label{r_2}
x^2_2=\frac{x^2_1g_m - 1}{g_m-1}.
\end{equation}

The value of $g_m$ can be determined from the relation found in \cite{Ott},
\begin{equation}\label{eq_gamma_gmax}
\Gamma=a_{1}(\kappa)+a_{2}(\kappa)g_m+a_{3}(\kappa)g_m^{2},
\end{equation}
where the $\kappa$-dependence of the parameters $a_{1}$, $a_{2}$, and $a_{3}$ are providing by using a second degree polynomial
\begin{equation}
a_{\xi}(\kappa)=c_{1}^{(\xi)}+c_{2}^{(\xi)}\kappa+c_{3}^{(\xi)}\kappa^{2},\,\,\xi=1,\,2,\,3\,. \nonumber
\end{equation}

For the case of a two-dimensional Yukawa fluid, the values of the dimensionless parameters $c_{1}^{(\xi)}$, $c_{2}^{(\xi)}$, and $c_{3}^{(\xi)}$ were are found by solving a system of nine cases of Equation (\ref{eq_gamma_gmax}) written for states with $\Gamma = 20$, $50$, and $100$, and $\kappa = 1$, $1.5$, and $2$. The solution was also carried out using the modified Newton method; the error of the numerical solution of this system of equations in this case did not exceed $0.03$ \%. The values for $g_m$ were collected from MD simulation data. As a result of solving this system, it was obtained that: $c_{1}^{(1)}=-248.56$, $c_{2}^{(1)}=369.596$, and $c_{3}^{(1)}=-126.792$; $c_{1}^{(2)}=272.541$, $c_{2}^{(2)}=-427.101$, and $c_{3}^{(2)}=137.019$; and $c_{1}^{(3)}=-58.808$, $c_{2}^{(3)}=112.002$, and $c_{3}^{(3)}=-28.326$. The relation (\ref{eq_gamma_gmax}) is actually the equation of state for the equilibrium liquid phase of the two-dimensional Yukawa system, from which we find
\begin{equation}\label{eq_res}
g_m=\frac{-a_{2}(\kappa)+\sqrt{a_{2}^{2}(\kappa)-4a_{1}(\kappa)a_{3}(\kappa)+4a_{3}(\kappa)\Gamma}}{2a_{3}(\kappa)}.
\end{equation}

Thus, the value of $x_2$ for a particular $(\Gamma, \; \kappa)$-state is determined by solving the system of Equations (\ref{eq_Gamma_r1}), (\ref{r_2}), and (\ref{eq_res}).

The RDF is included in microscopic expressions for many physical characteristics. Thus, the excess internal energy $U_{\rm ex}$ of a two-dimensional Yukawa fluid (in units of $k_BT$) is~\citep{HamaguchiJCP, Hartmann, KhrapakJCP, KhrapakPoP2016, Filippov}
\begin{equation}
U_{\rm ex} = \Gamma\int_0^\infty\ \exp(-\kappa x)g(x)\,dx.
\label{U_yukawa}
\end{equation}

Taking into account the approximation (\ref{eq_gr}), from the expression (\ref{U_yukawa}) we obtain
\begin{equation}
U_{\rm ex}^{(2\rm st)} = \frac{\Gamma g_m}{\kappa}\biggl[\exp(-\kappa x_1) - \frac{g_m-1}{g_m}\exp(-\kappa x_2)\biggr]. \label{U_2gr}
\end{equation}

Further, for the excess internal pressure $P_{\rm ex}$ of a two-dimensional Yukawa fluid (in units of $\rho k_BT$), we have: \citep{HamaguchiJCP, Hartmann, KhrapakJCP, KhrapakPoP2016, Filippov}
\begin{equation}
P_{\rm ex} = \frac{\Gamma}{2}\int_0^\infty\ \bigl(\kappa x+1\bigr)\exp(-\kappa x)g(x)\,dx\,.
\label{P_yukawa}
\end{equation}

Hence, taking into account the approximation (\ref{eq_gr}), we find
\begin{equation}
P_{\rm ex}^{(2\rm st)} = \frac{\Gamma g_m}{2\kappa}\biggl[\frac{\kappa x_1+2}{\exp(-\kappa x_1)} - \frac{g_m-1}{g_m}\frac{\kappa x_2+2}{\exp(\kappa x_2)}\biggr]. \label{P_2gr}
\end{equation}

The microscopic expression for the excess entropy $S_{ex2}$ in units of $\rho k_B$ in the two-particle approximation does not explicitly contain the interaction potential $u(r)$, and for a two-dimensional isotropic system has the form~\cite{SExcess}:
\begin{equation}
S_{\rm ex 2}=-\int_0^\infty[g(x)\ln g(x)+1-g(x)]x\,dx.
\label{S_2gr}
\end{equation}

Then, within the approximation (\ref{eq_gr}), from the expression (\ref{S_2gr}), we obtain
\begin{equation}
S_{\rm ex 2}^{(2\rm st)}=-\frac{1}{2(g_m-1)}\left[x_1^2g_m\ln g_m-\left(g_m\ln g_m+1-g_m\right)\right].
\label{S_2gr+2st}
\end{equation}

Knowing the RDF using the QLCA model, for a two-dimensional Yukawa fluid, one can calculate the longitudinal sound velocity $c_L$ (in units of thermal velocity\linebreak   $v_{\rm{th}}=\sqrt{k_BT/m}$) \cite{KhrapakPoP2016}:
\begin{equation}
c_L^2 =  \frac{\Gamma}{8}\int_0^\infty\bigl(3(\kappa x)^2 + 5\kappa x+5\bigr)\exp(-\kappa x)g(x)\,dx.
\label{cL}
\end{equation}

Within the framework of the approximation (\ref{eq_gr}), from the expression (\ref{cL}), we obtain
\begin{equation}
\begin{gathered}
c_L^{2(2\rm st)} = \frac{\Gamma g_m}{8\kappa}\biggl[\frac{3(\kappa x_1)^2+11\kappa x_1+16}{\exp(\kappa x_1)} \\
- \left(\frac{g_m-1}{g_m}\right)\frac{3(\kappa x_2)^2+11\kappa x_2+16}{\exp(\kappa x_2)}\biggr].
\label{cL+2st}
\end{gathered}
\end{equation}
As can be seen from the expressions (\ref{U_2gr}), (\ref{P_2gr}), (\ref{S_2gr+2st}), and (\ref{cL+2st}), the values $U_{ex}$, $P_ {ex}$, $S_{ex2}$, and $c_L$ can be directly calculated for a given ($\Gamma,\; \kappa$)-state. On the other hand, if the true function $g(x)$ is known, then the quantities can be estimated using microscopic expressions (\ref{U_yukawa}), (\ref{P_yukawa}), (\ref{S_2gr}), and (\ref{cL}).

The results of the numerical calculations of the reduced excess internal energy $U_{ex}$, the reduced excess internal pressure $P_{ex}$, and the reduced excess entropy $S_{ex2}$ performed within the~(\ref{eq_gr}) approximation for $g (x)$, as well as using the true $g(x)$ from the MD simulation, are presented in Table~ \ref{tableI}. This table also shows the relative correspondences between the theoretical results and simulation results. For most ($\Gamma,\; \kappa$)-states, the differences between the theoretical results and simulation data for $U_{ex}$ and $P_{ex}$ do not exceed $2$ \%. The largest discrepancies corresponding to $3.818$ \% for $U_{ex}$ and $2.912$ \% for $P_{ex}$ are observed for states with maximum $\kappa=2$, which can be explained by violation of the charge neutrality condition (\ref {charge_n1}) in the case of states with this value of the screening parameter $\kappa$. Further, the entropy $S_{ex2}$ is very structure sensitive. This may explain the weak agreement between the results of theoretical calculations and the data of the MD simulation.

\begin{table*}[tbp]
\caption{Reduced excess internal energy $U_{\rm ex}$, reduced excess internal pressure $P_{\rm ex}$, and reduced excess entropy $S_{\rm ex 2}$ of a 2D Yukawa fluid found using $g(x)$ from MD simulations. The same quantities ($U_{\rm ex}^{(2\rm st)}$, $P_{\rm ex}^{(2\rm st)}$, and $S_{\rm ex 2}^{ (2\rm st)}$) are calculated based on the expressions (\ref{eq_gr}), (\ref{P_2gr}),  and (\ref{S_2gr+2st}). The relative correspondences $\delta_{U_{\rm ex}}$ , $\delta_{P_{\rm ex}}$, and $\delta_{S_{\rm ex 2}}$ of these quantities in \% between theoretical results and simulation results.}

\label{tableI}
\newcolumntype{C}{>{\centering\arraybackslash}X}
\begin{tabular}{c c c c c c c c c c c}
\toprule
\boldmath{$\kappa$}\hspace{0.3cm} & \boldmath{$\Gamma$}\hspace{0.3cm} & \hspace{0.3cm}\boldmath{$U_{\rm ex}$}\hspace{0.3cm} & \boldmath{$U_{\rm ex}^{(2\rm st)}$}\hspace{0.3cm} & \boldmath{$\delta_{U_{\rm ex}}$}  \hspace{0.3cm} & \boldmath{$P_{\rm ex}$}\hspace{0.3cm} & \boldmath{$P_{\rm ex}^{(2\rm st)}$}\hspace{0.3cm} &\boldmath{ $\delta_{P_{\rm ex}}$} \hspace{0.3cm} & \boldmath{$S_{\rm ex 2}$}\hspace{0.3cm} & \boldmath{$S_{\rm ex 2}^{(2\rm st)}$}\hspace{0.3cm} & \boldmath{$\delta_{S_{\rm ex 2}}$} \\
\hline
1\hspace{0.3cm} & 20\hspace{0.3cm} & \hspace{0.3cm}6.762\hspace{0.3cm} & 6.767\hspace{0.3cm} & 0.072\hspace{0.3cm} & 10.554\hspace{0.3cm} & 10.536\hspace{0.3cm} & 0.165\hspace{0.3cm} & $-$0.643\hspace{0.3cm} & $-$0.782\hspace{0.3cm} & 21.576 \\
1\hspace{0.3cm} & 50\hspace{0.3cm} & \hspace{0.3cm}15.901\hspace{0.3cm} & 15.951\hspace{0.3cm} & 0.319\hspace{0.3cm} & 25.518\hspace{0.3cm} & 25.461\hspace{0.3cm} & 0.225\hspace{0.3cm} & $-$1.111\hspace{0.3cm} & $-$1.068\hspace{0.3cm} & 3.883 \\
1\hspace{0.3cm} & 100\hspace{0.3cm} & \hspace{0.3cm}30.943\hspace{0.3cm} & 30.641\hspace{0.3cm} & 0.977\hspace{0.3cm} & 50.294\hspace{0.3cm} & 49.719\hspace{0.3cm} & 1.142\hspace{0.3cm} & $-$1.803\hspace{0.3cm} & $-$1.348\hspace{0.3cm} & 25.226 \\
\hline
1.5\hspace{0.3cm} & 20\hspace{0.3cm} & \hspace{0.3cm}2.732\hspace{0.3cm} & 2.708\hspace{0.3cm} & 0.871\hspace{0.3cm} & 4.958\hspace{0.3cm} & 4.919\hspace{0.3cm} & 0.797\hspace{0.3cm} & $-$0.542\hspace{0.3cm} & $-$0.679\hspace{0.3cm} & 25.377 \\
1.5\hspace{0.3cm} & 50\hspace{0.3cm} & \hspace{0.3cm}5.963\hspace{0.3cm} & 6.068\hspace{0.3cm} & 1.751\hspace{0.3cm} & 11.403\hspace{0.3cm} & 11.470\hspace{0.3cm} & 0.591\hspace{0.3cm} & $-$0.897\hspace{0.3cm} & $-$0.941 \hspace{0.3cm} & 4.927\\
1.5\hspace{0.3cm} & 100\hspace{0.3cm} & \hspace{0.3cm}11.165\hspace{0.3cm} & 11.266\hspace{0.3cm} & 0.910\hspace{0.3cm} & 21.912\hspace{0.3cm} & 21.853\hspace{0.3cm} & 0.267\hspace{0.3cm} & $-$1.402\hspace{0.3cm} & $-$1.190\hspace{0.3cm} & 15.117 \\
\hline
2\hspace{0.3cm} & 20\hspace{0.3cm} & \hspace{0.3cm}1.336\hspace{0.3cm} & 1.285\hspace{0.3cm} & 3.818\hspace{0.3cm} & 2.697\hspace{0.3cm} & 2.619\hspace{0.3cm} & 2.912\hspace{0.3cm} & $-$0.454\hspace{0.3cm} & $-$0.581\hspace{0.3cm} & 27.931 \\
2\hspace{0.3cm} & 50\hspace{0.3cm} & \hspace{0.3cm}2.647\hspace{0.3cm} & 2.699\hspace{0.3cm} & 1.982\hspace{0.3cm} & 5.766\hspace{0.3cm} & 5.823\hspace{0.3cm} & 0.986\hspace{0.3cm} & $-$0.716\hspace{0.3cm} & $-$0.808\hspace{0.3cm} & 12.893 \\
2\hspace{0.3cm} & 100\hspace{0.3cm} & \hspace{0.3cm}4.665\hspace{0.3cm} & 4.815\hspace{0.3cm} & 3.225\hspace{0.3cm} & 10.601\hspace{0.3cm} & 10.756\hspace{0.3cm} & 1.468\hspace{0.3cm} & $-$1.053\hspace{0.3cm} & $-$1.026\hspace{0.3cm} & 2.562 \\
\end{tabular}
\end{table*}

\begin{figure}[h]
\begin{center}
\includegraphics[width=7.0cm]{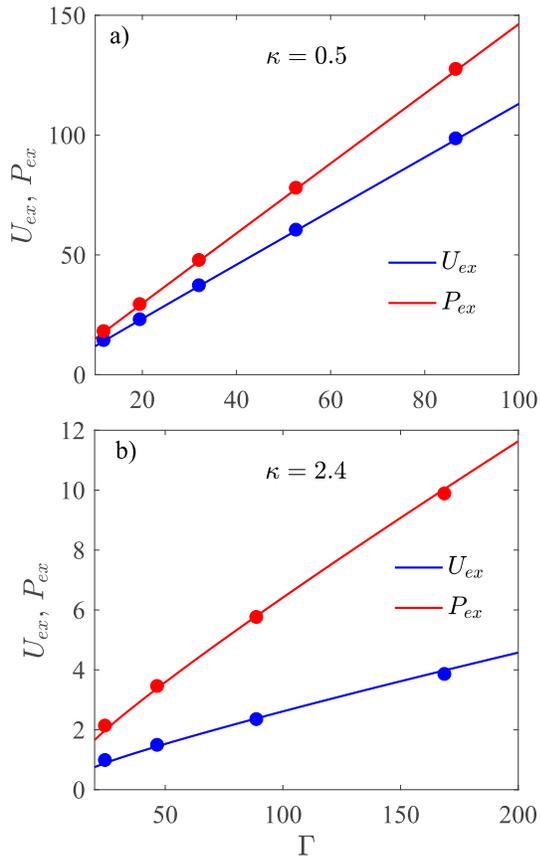}
\end{center}
\caption{Dependence of the reduced excess internal energy $U_{ex}$ and the reduced excess internal pressure $P_{ex}$ on the coupled parameter $\Gamma$ for the values of the screening parameter $\kappa=0.5$ (\textbf{a}) and $\kappa=2.4$ (\textbf{b}), which are constructed using expressions \eqref{U_2gr} and \eqref{P_2gr}. Symbols show data from work~\cite{Yurchenko}.}\label{figure}
\end{figure}

Figure \ref{figure} shows the dependence of the reduced excess internal energy $U_{ex}$ and reduced excess internal pressure $P_{ex}$ on the coupling parameter $\Gamma$ for the values of the screening parameter $\kappa$ outside the range that was used to construct a two-step approximation for RDF \eqref{eq_gr}. Here, we also show the data from the work \cite{Yurchenko}, in which the energy and pressure values of the two-dimensional Yukawa fluid were calculated in a wide range of changes in the value of the $\Gamma$ and $\kappa$. It can be seen that, even at these $\kappa$ values, the analytical formulas obtained in this work for the direct calculation of reduced excess internal energy and pressure of two-dimensional Yukawa fluids are in good agreement with the simulation data.

Table \ref{tableII} shows the results of the numerical calculations of the reduced longitudinal sound velocity $c_L$, performed within the approximation~(\ref{eq_gr}) for $g(x)$, as well as using the true $g(x)$ from MD modeling. This table also shows the relative correspondences between theoretical results and simulation results. The greatest discrepancy is observed for the state of the two-dimensional Yukawa liquid with $\Gamma=20$ and $\kappa=2$, which is the closest of all those considered to the gas phase. This feature is related to the fact that the approximation of a quasi-localized charge better describes the collective properties of a Yukawa liquid with states close to crystalline, i.e., with large $\Gamma$ and small $\kappa$ \cite{Kalman}.

\begin{table}[h]
\caption{Longitudinal sound velocity in units of thermal velocity $c_L$ of a 2D Yukawa fluid, found using $g(x)$ from MD simulations. The same value $c_L^{(2\rm st)}$ calculated based on the expression (\ref{cL+2st}). The relative correspondences $\delta_{c_L}$ of this value in \% between theoretical and simulation results are also provided.}
\label{tableII}
\begin{tabular}{c c  c c c}

\toprule
\boldmath{$\kappa$}\hspace{0.3cm} & \boldmath{$\Gamma$}\hspace{0.3cm} & \hspace{0.3cm}\boldmath{$c_L$}\hspace{0.3cm} & \boldmath{$c_L^{(2\rm st)}$}\hspace{0.3cm} & \boldmath{$\delta_{c_L}$} \\
\hline
1\hspace{0.3cm} & 20\hspace{0.3cm} & \hspace{0.3cm}5.196\hspace{0.3cm} & 5.189\hspace{0.3cm} & 0.139 \\
1\hspace{0.3cm} & 50\hspace{0.3cm} & \hspace{0.3cm}8.145\hspace{0.3cm} & 8.13\hspace{0.3cm} & 0.184 \\
1\hspace{0.3cm} & 100\hspace{0.3cm} & \hspace{0.3cm}11.475\hspace{0.3cm} & 11.422\hspace{0.3cm} & 0.465 \\
\hline
1.5\hspace{0.3cm} & 20\hspace{0.3cm} & \hspace{0.3cm}3.777\hspace{0.3cm} & 3.761\hspace{0.3cm} & 0.434 \\
1.5\hspace{0.3cm} & 50\hspace{0.3cm} & \hspace{0.3cm}5.831\hspace{0.3cm} & 5.827\hspace{0.3cm} & 0.077 \\
1.5\hspace{0.3cm} & 100\hspace{0.3cm} & \hspace{0.3cm}8.156\hspace{0.3cm} & 8.122\hspace{0.3cm} & 0.422 \\
\hline
2\hspace{0.3cm} & 20\hspace{0.3cm} & \hspace{0.3cm}2.917\hspace{0.3cm} & 2.881\hspace{0.3cm} & 1.266 \\
2\hspace{0.3cm} & 50\hspace{0.3cm} & \hspace{0.3cm}4.391\hspace{0.3cm} & 4.392\hspace{0.3cm} & 0.030 \\
2\hspace{0.3cm} & 100\hspace{0.3cm} & \hspace{0.3cm}6.051\hspace{0.3cm} & 6.055\hspace{0.3cm} & 0.063 \\

\end{tabular}

\end{table}

In order to check the validity of the relation \eqref{cL+2st} in the case of values of the coupling and screening parameters outside the range used in this work when constructing the two-step approximation for the $g(x)$, we calculated the longitudinal sound velocity for the gamma--kappa states collected from \cite{KhrapakPoP2016}. The results are presented in Table \ref{tableIII}. It can be seen that, even at $\Gamma = 1033$ and $\kappa=3$, formula \eqref{cL+2st} provides a deviation of less than 4\% from the value of s calculated using the true $g(x)$.

\begin{table}[h]
\caption{The same as in Table 2, except for the third column; here are the data from work \cite{KhrapakPoP2016}.}
\label{tableIII}
\newcolumntype{C}{>{\centering\arraybackslash}X}
\begin{tabular}{c c  c c c}

\hline
\boldmath{$\kappa$}\hspace{0.3cm} & \boldmath{$\Gamma$}\hspace{0.3cm} & \hspace{0.3cm}\boldmath{$c_L$}\hspace{0.3cm} & \boldmath{$c_L^{(2\rm st)}$}\hspace{0.3cm} & \boldmath{$\delta_{c_L}$} \\
\hline
1\hspace{0.3cm} & 163\hspace{0.3cm} & \hspace{0.3cm}14.62\hspace{0.3cm} & 14.51\hspace{0.3cm} & 0.75 \\
2\hspace{0.3cm} & 362\hspace{0.3cm} & \hspace{0.3cm}11.25\hspace{0.3cm} & 11.02\hspace{0.3cm} & 2.04 \\
3\hspace{0.3cm} & 1033\hspace{0.3cm} & \hspace{0.3cm}10.23\hspace{0.3cm} & 9.87\hspace{0.3cm} & 3.52 \\
\hline

\end{tabular}
\end{table}
The results of this work indicate the following. The thermodynamic characteristics and the sound velocity for a two-dimensional Yukawa fluid can be theoretically calculated from microscopic expressions, where the characteristic parameters of the Yukawa system (coupling parameter and screening parameter) are used as input parameters. The two-step approximation proposed in this work for the RDF provides a good agreement with the simulation results for such quantities as the internal energy, internal pressure, and longitudinal sound velocity. If we compare the results for the two-dimensional Yukawa fluid with the three-dimensional case \cite{Fairushin}, we can see that the obtained analytical \mbox{expressions \eqref{U_2gr} and \eqref{P_2gr}} are simpler and that, at the same time, they provide the same accuracy when reproducing the simulations results. In addition, as in the case of the three-dimensional Yukawa fluid in the two-dimensional case, to correctly calculate the excess entropy, it is necessary to use a more accurate model for $g(r)$.

\vskip 0.5 cm

\acknowledgments This work was supported by the Russian Science Foundation
(Project No. 19-12-00022). The authors are grateful to S.A. Khrapak for helpful discussions.

\end{document}